\documentclass[twocolumn,showpacs,preprintnumbers,amsmath,amssymb,aps,prl,superscriptaddress]{revtex4-1}
\usepackage{epsfig}
\usepackage{graphicx}% Include figure files
\usepackage{dcolumn}% Align table columns on decimal point
\usepackage{bm}% bold math
\usepackage{color} % \color{red} \color{black}

\begin{document}

\title{
Existence of Heavy Fermions in the Antiferromagnetic Phase of CeIn$_3$
}
\author{Takuya Iizuka}
\email{takuizk@ims.ac.jp}
\author{Takafumi Mizuno}
\affiliation{School of Physical Sciences, The Graduate University for Advanced Studies (SOKENDAI), Okazaki 444-8585, Japan}
\author{Byeong Hun Min}
\author{Yong-Seung Kwon}
\affiliation{Department of Emerging Materials Science, Daegu Gyeongbuk Institute of Science and Technology (DGIST), Daegu 711-873, Republic of Korea}
\author{Shin-ichi Kimura}
\email{kimura@ims.ac.jp}
\affiliation{UVSOR Facility, Institute for Molecular Science, Okazaki 444-8585, Japan}
\affiliation{School of Physical Sciences, The Graduate University for Advanced Studies (SOKENDAI), Okazaki 444-8585, Japan}
\date{\today}
\begin{abstract}
We report the pressure-dependent optical conductivity spectra of a heavy fermion (HF) compound CeIn$_3$ below the N\'eel temperature of 10~K to investigate the existence of the HF state in the antiferromagnetic (AFM) phase.
The peak due to the interband transition in the hybridization gap between the conduction band and nearly localized $4f$ states ($c$-$f$ hybridization) appears at the photon energy of about 20~meV not only in the HF regime but also in the AFM regime.
Both the energy and intensity of the $c$-$f$ hybridization peak continuously increase with the application of pressure from the AFM to the HF regime.
This result suggests that the $c$-$f$ hybridization, as well as the heavy fermions, exists even in the AFM phase of CeIn$_3$.
\end{abstract}
%
%\kword{CeIn$_3$, optical conductivity, electronic structure, quantum critical point}
\pacs{71.27.+a, 78.20.-e}% PACS, the Physics and Astronomy
                             % Classification Scheme.
%%%%%%%%%%%%%%%%%%%%%%%%%%%%%%%%%%%%%%%%%%%%%%%%%%%%%%%%%%%%
%\begin{document}
\maketitle
%
%%%%%%%%%%%%%%%%%%%%%%%%%%%%%%%%%%%%%%%%%%%%%%%%%%%%%%%%%%%%
%\section{Introduction}
%
Recently, physics at the quantum critical point (QCP), which is the border between local magnetism and itinerant paramagnetism at zero temperature, has become one of the main topics in the condensed-matter field because new quantum properties such as non-BCS superconductivity appear in the vicinity of the QCP. 
The ground state of rare-earth intermetallic compounds, namely, heavy-fermion (HF) materials, changes between the local magnetic and itinerant nonmagnetic states through external perturbation by such factors as pressure and magnetic field~\cite{Gegenwart2008}. 
The QCP appears owing to the energy balance between the local magnetic state based on the Ruderman-Kittel-Kasuya-Yoshida (RKKY) interaction and the itinerant HF state due to the Kondo effect. 
In the itinerant HF regime, the conduction band hybridizes with the nearly local $4f$ state, so that a large Fermi surface as well as the hybridization band between them, namely, the $c$-$f$ hybridization band, is realized. 
In the case of a magnetic regime, on the other hand, two theoretical scenarios have been proposed. 
One is the spin-density wave (SDW) scenario based on spin fluctuation, in which large Fermi surfaces due to $c$-$f$ hybridization remain even in magnetically ordered states. 
The other is the Kondo breakdown (KBD) scenario, in which the $c$-$f$ hybridization state disappears in the magnetic state and only small Fermi surfaces due to conduction electrons appear~\cite{Coleman2001}.
Many controversies for these scenarios have been performed so far, but the conclusion has not been obtained yet.

Experimental studies have supported both scenarios as follows.
A neutron scattering experiment on Ce(Ru$_{0.97}$Rh$_{0.03}$)$_2$Si$_2$ has found itinerant antiferromagnetic (AFM) behavior at the QCP, supporting the SDW scenario~\cite{Kadowaki2006}.
On the other hand, Hall effect measurements of YbRh$_2$Si$_2$ have shown a marked increase in the carrier density at the QCP, supporting the KBD scenario~\cite{Paschen2004,Friedemann2010},
while pressure-dependent de Haas-van Alphen (dHvA) effect measurements of CeRhIn$_5$ and CeIn$_3$ have also indicated a discontinuous change in the Fermi surface across the critical pressure~\cite{Shishido2005,Settai2005}.
The drastic Fermi surface reconstruction seems to be explained by the KBD scenario.
Recent theoretical works, however, have shown that the valence transition is the main origin of these discontinuous changes of dHvA~\cite{Watanabe2008,Watanabe2010}.
To clarify the change in the electronic structure across the critical pressure, it is therefore necessary to investigate the electronic structure at low temperatures as a function of $c$-$f$ hybridization intensity.

Optical measurements are highly useful means of clarifying the electronic structure at the the critical pressure.
For example, the $c$-$f$ hybridization state can be directly revealed by optical conductivity [$\sigma(\omega)$] measurements in the far-infrared region~\cite{Kimura2006-1,Kimura2011} as well as angle-resolved photoemission spectroscopy (ARPES)~\cite{Im2008,Klein2011}.
If such optical measurements are performed from the local regime to the itinerant regime controlled by the pressure or magnetic field, any change in the electronic structure across the QCP is revealed.
Since ARPES cannot be performed in a pressure cell or a high magnetic field, $\sigma(\omega)$ measurements are one of the most realistic methods of detecting changes in the electronic structure due to the pressure and/or magnetic field.

In this Letter, we describe the pressure-dependent electronic structure as well as the $c$-$f$ hybridization state obtained by far-infrared reflectivity [$R(\omega)$] and $\sigma(\omega)$ measurements of CeIn$_3$ under pressure. 
CeIn$_3$ has an AFM ground state with a N\'eel temperature $T_{\rm N}$ of 10 K. 
With the application of pressure, $T_{\rm N}$ monotonically decreases and disappears at a critical pressure of approximately 2.6 GPa~\cite{Mathur1998,Knebel2001,Grosche2001}.
We observed that the $c$-$f$ hybridization gap appears not only in the HF state but also in the AFM state, and both the energy and intensity of the $\sigma(\omega)$ peak due to the $c$-$f$ hybridization band continuously increase with the application of pressure.
Our observations suggest that the electronic structure of CeIn$_3$ in the AFM phase can be explained by the SDW scenario because the $c$-$f$ hybridization state exists even in the AFM phase.

%%%%%%%%%%%%%%%%%%%%%%%%%%%%%%%%%%%%%%%%%%%%%%%%%%
%\section{Experimental}
%
CeIn$_3$ samples were synthesized by an arc melting method, and then annealed at 900~$^\circ$C for 3~weeks inside an evacuated quartz tube~\cite{Lee2008}. 
Measurements of the pressure-dependent $R(\omega)$ spectra in the far-infrared region [120~cm$^{-1}$ (14~meV) to 400~cm$^{-1}$ (50~meV)] were performed at the terahertz microscopy end-station of beamline 6B of UVSOR-II at the Institute for Molecular Science~\cite{Kimura2006-2}.
A gas membrane-type diamond anvil cell (DAC) [WCM-7(B) Diacell OptiDAC-LT, easyLab Technologies Ltd.] with a culet size of 1 mm was employed to apply high pressure to the samples. 
The DAC was cooled to 5.6~K using a liquid helium-flow-type cryostat (ST-100, Janis Research Co.).
Samples for measurement were shaped to dimensions of 0.3~mm diameter and 0.02~mm thickness from a sample block. 
The sample and a gold film as a reference material were set inside a pinhole (0.6~mm in diameter) of the gasket (stainless steel) of the DAC. 
KBr powder and Apiezon-N grease were used as pressure media. 
Ruby powder was set at several points around the sample to monitor pressure, and the applied pressure was evaluated using a ruby fluorescence method~\cite{Ragen1992}.
The pressure distribution was evaluated to be less than about $\pm$0.1~GPa from the broadening of the ruby fluorescence spectrum.
We also measured the temperature-dependent $R(\omega)$ spectra in a wide energy range from 3~meV to 30~eV using the synchrotron and conventional apparatuses~\cite{Iizuka2010}, and $\sigma(\omega)$ spectra were obtained by means of the Kramers-Kronig (K-K) analysis~\cite{DresselGruner}.

%%%%%%%%%%%%%%%%%%%%%%%%%%%%%%%%%%%%%%%%%%%%%%%%%
%\section{Results and Discussion}
%
%%%%%%%%%%%%%%  FIG.1. s(w) of CeIn$_3$  %%%%%%%%%%%%%%%%%%%%
\begin{figure}[b]
\begin{center}
\includegraphics[width=0.45\textwidth]{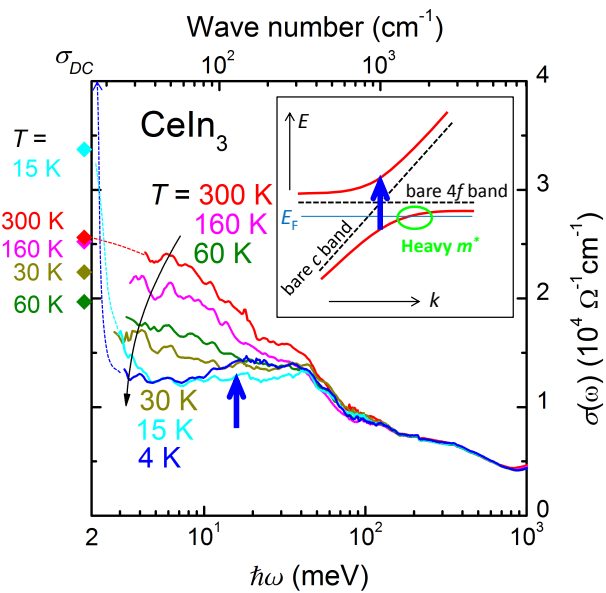}
\end{center}
\caption{
(Color online)
Temperature-dependent optical conductivity [$\sigma(\omega)$] spectra of CeIn$_3$ at ambient pressure. 
The solid diamonds show the direct current conductivities ($\sigma_{DC}$) reported in ref. 15.
Extrapolations of the $\sigma(\omega)$ spectra to $\sigma_{DC}$ values at 300, 15 and 4~K are depicted by dotted lines.
($\sigma_{DC}$ of $1.8\cdot10^5$~$\Omega^{-1}{\rm cm}^{-1}$ at 4~K is out of range.)
(Inset) Schematic figure of the dispersion curve of the $c$-$f$ hybridization state based on the periodic Anderson model.~\cite{Hewson}
Dispersion curves of bare $4f$ and conduction ($c$) bands are indicated by dashed lines and the $c$-$f$ hybridization bands are solid lines.
The $\sigma(\omega)$ peak and narrow Drude peak originate from the interband transition indicated by an arrow and the heavy mass ($m^*$) band at the Fermi level, respectively.
}
\label{OC}
\end{figure}
%%%%%%%%%%%%%%%%%%%%%%%%%%%%%%%%%%%%%%%%%%%%%%%%%
We describe the temperature-dependent $\sigma(\omega)$ spectra of CeIn$_3$ to clarify the electronic structure at ambient pressure (Fig.~\ref{OC}).
There is no conspicuous spectral change with temperature in the photon energy region above 40~meV, because the $\sigma(\omega)$ spectra originate from the optical transition from the occupied electronic state just below the Fermi level ($E_{\rm F}$) to the unoccupied state mainly consisting of the Ce $4f_{5/2}$ and $4f_{7/2}$ states predicted by a band structure calculation~\cite{Kimura2009}.
The main temperature dependence appears below 40~meV. 
Above 60~K, the $\sigma(\omega)$ intensity below 40~meV gradually decreases with cooling and the extrapolations to the zero photon energy of the $\sigma(\omega)$ spectra are consistent with the direct current conductivity ($\sigma_{DC}$)~\cite{Knebel2001}. 
This suggests that the decrease in $\sigma_{DC}$ (increase in electrical resistivity) originates from the decrease in spectral weight near $E_{\rm F}$. 
Below 30~K, the spectral shape changes from the ordinary metallic Drude one to the HF one with a peak due to the $c$-$f$ hybridization gap at about 20~meV and a narrow Drude peak due to quasiparticles below 5~meV~\cite{Kimura2006-1,Kimura2011,Iizuka2010}.
The change in the electronic structure based on the periodic Anderson model~\cite{Hewson} is schematically depicted in the inset of Fig.~\ref{OC}.
The bare $4f$ and conduction ($c$) bands at high temperatures (dashed lines) hybridize with each other at low temperatures (solid lines), and then a $c$-$f$ hybridization gap appears.
A similar peak (shoulder) has been observed in HF materials such as CeRu$_4$Sb$_{12}$~\cite{Dordevic2001}, CeCu$_6$~\cite{Marabelli1990}, CePd$_3$~\cite{Kimura2009}, CeNiGe$_2$~\cite{Kwon2006}, and YbIr$_2$Si$_2$~\cite{Iizuka2010}.
Among these materials, CeRu$_4$Sb$_{12}$ and CeCu$_6$ are located in the vicinity of the QCP, and YbIr$_2$Si$_2$ and CePd$_3$ are located in the itinerant regime.
In the case of the title compound CeIn$_3$ with the AFM ground state, a similar $c$-$f$ hybridization electronic structure is realized.
This situation is the same as that of CeNiGe$_2$.

%%%%%%%%%%%%%%  FIG.2. Optical conductivity spectra under pressure  %%%%%%%%%%%%%%%%%%%%
\begin{figure}[b]
\begin{center}
\includegraphics[width=0.45\textwidth]{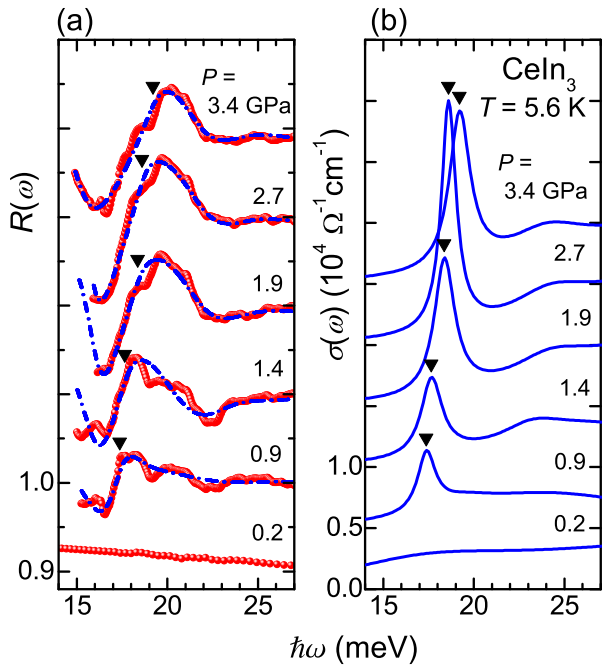}
\end{center}
\caption{
(Color online)
(a) Pressure dependence of the reflectivity [$R(\omega)$] spectrum (solid circles) of CeIn$_3$ after a correction procedure (see text) and Drude-Lorentz fitting results (dotted-dashed lines) in the photon energy $\hbar\omega$ range of 14--27~meV at 5.6~K.
The spectra are shifted by 0.1 for clarity.
(b) Optical conductivity [$\sigma(\omega)$] spectra derived from the Drude-Lorentz fitting of the [$R(\omega)$] spectra in (a).
The Drude component was subtracted.
These spectra are offset by $5\times10^3$~$\Omega^{-1}{\rm cm}^{-1}$ for clarity.
The solid triangles show the center energies of the fitted Lorentzian functions.
}
\label{Pdep}
\end{figure}
%%%%%%%%%%%%%%%%%%%%%%%%%%%%%%%%%%%%%%%%%%%%%%%%%
To investigate the pressure effect of the $c$-$f$ hybridization gap, pressure-dependent reflectivity [$R(\omega, P)$] spectra were measured at 5.6~K. 
Because of diffraction and other extrinsic effects, it is difficult to obtain the absolute value of $R(\omega)$~\cite{Sacchetti2007}.
Therefore, we calculated the absolute $R(\omega)$ spectra in the DAC using the following method.
First, the expected reflectivity spectrum [$R_{ideal}(\omega)$], in which the sample surface is in direct contact with a diamond (refractive index: 2.4) of the DAC, was calculated from the optical constants derived from the K-K analysis for the $R(\omega)$ spectrum at ambient pressure.
Second, we regarded the reflectivity spectrum in the DAC [$R_{DAC}(\omega, P)$] at the lowest pressure of 0.2~GPa to be the same as $R_{ideal}(\omega)$, because no conspicuous change in the physical properties was reported. 
The pressure-induced spectral change relative to that at 0.2~GPa [$R_{DAC}(\omega,P)/R_{DAC}(\omega,P=0.2~{\rm GPa})$] was measured and the corrected $R(\omega,P)$ spectra were obtained by multiplication of the $R_{ideal}(\omega)$ spectrum [$R(\omega,P)=R_{ideal}(\omega) \times {R_{DAC}(\omega,P)/R_{DAC}(\omega,P=0.2~{\rm GPa})}$].

The obtained $R(\omega,P)$ spectra are shown in Fig.~\ref{Pdep}(a).
The $R(\omega,P = 0.2~{\rm GPa}$) spectrum, which is the same spectrum as $R_{ideal}(\omega)$, is almost flat, because the peak is not large.
With increasing pressure, however, a significant dispersive structure appears and the intensity increases.
To obtain the pressure-dependent optical conductivity [$\sigma(\omega,P)$] spectra from the $R(\omega,P)$ spectra, fitting using a Drude and some Lorentzian functions, namely, Drude-Lorentz fitting, was applied~\cite{DresselGruner}.
The fitting functions are as follows:
\[
R(\omega)=\left|\{1-\tilde{\varepsilon}(\omega)^{1/2}\}/\{1+\tilde{\varepsilon}(\omega)^{1/2}\}\right|^2, 
\]
\[
\tilde{\varepsilon}(\omega)=\varepsilon_{\infty}+\sum_j\omega_{pj}^2/(\omega_{0j}^2-\omega^2-i\omega\Gamma_j) ,
\]
where $\tilde{\varepsilon}(\omega)$ is a complex dielectric function, and $\varepsilon_{\infty}$ is the sum of $\varepsilon_1(\omega)$ above the measured energy region and was set as 1.
The variables $\omega_{pj}$, $\omega_{0j}$, and $\Gamma_j$ are the plasma frequency, center frequency (for a Drude: $\omega_{0j}=$~0), and scattering rate, respectively. 
The initial values of these parameters were obtained from the fitting of the $\sigma(\omega)$ spectra at ambient pressure at 4~K, and the parameters for the higher-energy region were assumed not to change with pressure.
The parameters for the Drude part were fixed, because the change of the Drude weight cannot be recognized from the spectra owing to the limited spectral region.
Using the fitting parameters, the $\sigma(\omega)$ spectra can be calculated using the following function~\cite{DresselGruner}:
\[
\sigma(\omega)=Re[\omega\{\tilde{\varepsilon}(\omega)-1\}/(4\pi i)].
\]

Figure~\ref{Pdep}(b) shows the $\sigma(\omega,P)$ spectra, in which the Drude component was subtracted, obtained using the above procedure. 
At a pressure of 0.2~GPa, the $\sigma(\omega)$ spectrum is almost flat but a broad peak is slightly visible at approximately 17~meV.
At ambient pressure, as shown in Fig.~\ref{OC}, the $\sigma(\omega)$ spectrum at 4~K has a broad peak at approximately 20~meV owing to the $c$-$f$ hybridization gap, but the intensity is smaller than that of other HF materials. 
Figure~\ref{Pdep}(b) clearly shows that a peak emerges at pressures above 0.9~GPa, shifts to the higher energy side, and increases with the application of pressure. 
Note that the $\sigma(\omega)$ peak is strikingly sharp at 2.7~GPa near the critical pressure.
This suggests that the flat bonding and antibonding bands of the $c$-$f$ hybridization state are located near $E_{\rm F}$ near the critical pressure and the interband transition between these bands appears in the $\sigma(\omega)$ spectrum.
This spectral feature of a peak at higher pressures is very similar to those of CeCu$_6$~\cite{Marabelli1990} and CeRu$_4$Sb$_{12}$~\cite{Dordevic2001}, which are located near the QCP, at ambient pressure.
The peaks of CeCu$_6$ and CeRu$_4$Sb$_{12}$ have been attributed to the optical transition between the bonding and antibonding states of the $c$-$f$ hybridization gap, as shown in the inset of Fig.~\ref{OC}. 
Therefore, the peak of CeIn$_3$ is considered to be the same optical transition in the $c$-$f$ hybridization gap. 
The peak shifts to the higher energy side with increasing pressure.
This is also consistent with the hybridization effect because the hybridization gap size, as well as the hybridization intensity, increases as the lattice constant decreases.

%%%%%%%%%%%%%%  FIG.3. Phase diagram  %%%%%%%%%%%%%%%%%%%%
\begin{figure}[b]
\begin{center}
\includegraphics[width=0.45\textwidth]{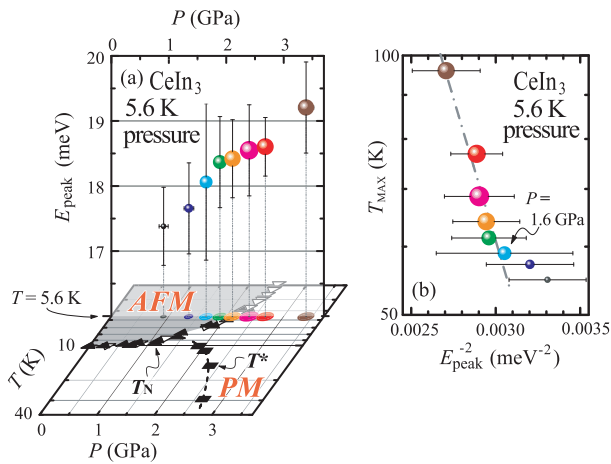}
\end{center}
\caption{
(Color online)
(a) Peak energies ($E_{peak}$) of the $\sigma(\omega)$ spectra in Fig.~\ref{Pdep}(b) as functions of pressure. 
The size of the marks denotes the intensity of the corresponding peak in the $\sigma(\omega)$ spectra. 
The pressure-dependent N\'eel temperature (solid and open triangles, $T_{\rm N}$) and valence transition temperature (solid squares, $T^*$) are also plotted at the bottom. (see ref. 28 for detail.) 
(b) The temperature of the maximum electrical resistivity ($T_{MAX}$)~\cite{Knebel2001} as a function of $E_{peak}^{-2}$ of the $\sigma(\omega)$ spectra. 
The linear dependence between logarithmic $T_{MAX}$ and $E_{peak}^{-2}$ is shown by the dotted-dashed line.
}
\label{GapPlot}
\end{figure}
%%%%%%%%%%%%%%%%%%%%%%%%%%%%%%%%%%%%%%%%%%%%%%%%%
The energy and intensity of the $\sigma(\omega)$ peaks in Fig.~\ref{Pdep}(b) are plotted by the position and size of marks, respectively, as functions of pressure in Fig.~\ref{GapPlot}(a). 
The figure shows that the $\sigma(\omega)$ peak monotonically shifts to the higher-energy side and grows with increasing pressure. 
The $\sigma(\omega)$ peak intensity rapidly increases at approximately 1.6~GPa (in the AFM phase), as shown in Fig.~\ref{Pdep}(b). 
Since the peak indicates the $c$-$f$ hybridization gap formation, the $c$-$f$ hybridization state appears even in the AFM phase.
In Fig.~\ref{GapPlot}(a), the pressure-dependent $T_{\rm N}$ and the crossover temperature $T^*$ between the localized and itinerant regimes of the $4f$ electrons observed in a previous nuclear quadrupole resonance (NQR) experiment are also plotted at the bottom~\cite{Kawasaki2001}.
%In the paramagnetic (PM) region, the valence transition was observed as the $T^*$ line in the previous NQR experiment. 
The observed pressure (1.6~GPa) of the $c$-$f$ hybridization gap at 5.6~K is roughly located on the extended line of $T^*$.
Therefore, the $T^*$ line can be extended from the PM phase to the AFM phase, further verifying the applicability of the SDW scenario depicted in ref.~1.

To confirm the origin of the $\sigma(\omega)$ peak in Fig.~\ref{GapPlot}(a), the relation of the peak energy and the pressure-dependent Kondo temperature ($T_{\rm K}$) was determined.
The relation of $T_{\rm K}$ and the effective hybridization ($\tilde{V}$) is 
\[
\ln k_B T_K \propto -\tilde{V}^{-2}D_c(E_F)^{-1},
\]
where $k_B$ is the Boltzmann constant and $D_c(E_F)$ is the density of states of the conduction band at $E_{\rm F}$~\cite{Hewson}.
According to the periodic Anderson model, the $\sigma(\omega)$ peak energy ($E_{peak}$) is proportional to $\tilde{V}$.
Since $T_{\rm K}$ is roughly proportional to the temperature of maximum electrical resistivity ($T_{MAX}$)~\cite{Knebel2001}, particularly in the case of measurements at a pressure that makes it possible to measure different hybridization strengths using the same sample, logarithmic $T_{MAX}$ should be proportional to $E_{peak}^{-2}$. 
As shown by the dotted-dashed line in Fig.~\ref{GapPlot}(b), the relation is satisfied in the pressure range above 1.6~GPa, which is in the AFM phase.
This plot suggests again that a $c$-$f$ hybridization state related to the Kondo effect exists even in the AFM phase, i.e., the electronic structure as well as the phase diagram of CeIn$_3$ can be explained by the SDW scenario.

%%%%%%%%%%%%%%%%%%%%%%%%%%%%%%
%\section*{Conclusion}
%
In summary, we measured the pressure-dependent reflectivity spectra of CeIn$_3$ at 5.6~K in the far-infrared region to investigate the change in the electronic structure from the local antiferromagnetism (AFM) to the itinerant heavy-fermion regime. 
A peak at approximately 18~meV clearly appears above 1.6~GPa even in the AFM phase.
Because the pressure dependence of the peak energy can be scaled to the temperature of maximum electrical resistivity, the observed peak is considered to originate from the hybridization between the conduction band and the $4f$ states ($c$-$f$ hybridization). 
This result implies that the $c$-$f$ hybridization, as well as the valence transition, appears even in the AFM phase described by the spin-density wave scenario.

%
%%%%%%%%%%%%%%%%%%%%%%%%%%%%%%
\section*{Acknowledgments}
We would like to thank UVSOR staff members for their technical support.
Part of this work was supported by the Use-of-UVSOR Facility Program (BL6B, 2007-2009) of the Institute for Molecular Science.
The work was partly supported by a Grant-in-Aid for Scientific Research (B) (Grant Nos.~18340110, 22340107) from JSPS of Japan and by the Nuclear R\&D Programs (2006-2002165) through the NRF funded by the Ministry of Education, Science and Technology of Republic of Korea.

%%%%%%%%%%%%%%%%%%%%%%%%%%%%%%

%
\end{document}